# Baker-Campbell-Hausdorff expansion in creation-only coupled cluster

Ramón Alain Miranda-Quintana

Department of Chemistry and Quantum Theory Project; University of Florida; Gainesville, FL, 32608, USA

Email: quintana@chem.ufl.edu

We study the properties of the exponential ansatz formed by strings of creation operators. We discuss the conditions under which we can recover a structure similar to traditional coupled cluster methods. We pay particular attention to the Baker-Campbell-Hausdorff expansion of these operators.

Exponential operators have a long history in theoretical chemistry, from Jastrow factors[1-3] to their use in orbital optimization.[4,5] But perhaps these operators are better known for their central role in the exponential coupled cluster (CC) ansatz.[6-10] In this case, the wavefunction is written as:

$$|\Psi_{CC}\rangle = \exp(\hat{T})|\Phi_0\rangle \qquad (1)$$

where the cluster operator, $\hat{T}$, is just a linear combination of excitation operators:

$$\hat{T} = \sum_\mu t_\mu \hat{\tau}_\mu \qquad (2)$$

$$\hat{\tau}_\mu = a^\dagger_{a_\mu} a^\dagger_{b_\mu} ... a^\dagger_{c_\mu} a_{i_\mu} a_{j_\mu} ... a_{k_\mu} \quad (3)$$

Operators $a_i^\dagger$ and $a_i$ create and annihilate a fermion in state $i$, respectively, and indices $i, j, ..., k$ ($a, b, ..., c$) refer to occupy (virtual) states in the reference $|\Phi_0\rangle$.

It is worth noting that the excitation operators possess some properties that make them particularly attractive, namely:

1- They commute with each other:

$$[\hat{\tau}_\mu, \hat{\tau}_\nu] = 0 \quad (4)$$

which means that we can write:

$$|\Psi_{CC}\rangle = \exp\left(\sum_\mu t_\mu \hat{\tau}_\mu\right)|\Phi_0\rangle = \prod_\mu \exp(t_\mu \hat{\tau}_\mu)|\Phi_0\rangle \quad (5)$$

2- They are square-zero operators:[11]

$$\hat{\tau}_\mu^2 = 0 \quad (6)$$

which means that we can also formulate the CC wavefunction using a product ansatz:

$$|\Psi_{CC}\rangle = \prod_\mu \exp(t_\mu \hat{\tau}_\mu)|\Phi_0\rangle = \prod_\mu (1 + t_\mu \hat{\tau}_\mu)|\Phi_0\rangle \quad (7)$$

Usually, the exponential formulation of CC is praised because it enforces size-extensivity.[5,12] However, here we will focus on another advantage, namely, the possibility that exponential operators give to work with the unliked formulation of CC. In this case, we use a similarity-transformed Hamiltonian, $\hat{H}^T$:

$$\hat{H}^T = \exp(-\hat{T})\hat{H}\exp(\hat{T}) \quad (8)$$

from where we can use the Baker-Campbell-Hausdorff (BCH) expansion:[13,14]

$$\exp(-\hat{T})\hat{H}\exp(\hat{T}) = \hat{H} + \sum_k \frac{1}{k!}\left[\hat{H},\hat{T}\right]_k \quad (9)$$

where:

$$\left[\hat{H},\hat{T}\right]_k = \left[...\left[\left[\hat{H},\hat{T}\right],\hat{T}\right]...,\hat{T}\right], k \text{ times} \quad (10)$$

In the form given in Eq. (9) it might appear that the BCH expansion complicates the problem at hand since it seems to result in an infinite sum of nested commutators. However, it can be shown that this expansion contains, at most, $4^{th}$ order commutators.[5] This provides a convenient way to obtain closed expressions to calculate the energy and cluster amplitudes.

In this Note we focus on exponential operators of linear combinations of products of creation operators, namely:

$$\exp(\hat{C})$$
$$\hat{C} = \sum_\mu c_\mu \hat{\chi}_\mu \quad (11)$$
$$\hat{\chi}_\mu = a^\dagger_{a_\mu} a^\dagger_{b_\mu} ... a^\dagger_{c_\mu}$$

These operators have been used to describe inter-geminal correlation in antisymmetrized geminal power (AGP) wavefunctions.[15] Additionally, they provide a way to generate particle number symmetry-broken wavefunctions and could be used in multi-reference CC studies.[7,16-19]

The first thing we want to check is whether a set of $\hat{\chi}_\mu$ operators satisfy the properties of the excitation operators that make them amenable to a CC formulation. First, it is easy to see that since two fermions must occupy different states, the $\hat{\chi}_\mu$'s will always be zero-square operators. The case of commutativity, on the other hand, is not so straightforward, since for an arbitrary pair of $\hat{\chi}_\mu$ operators to commute we can have at most one such operator formed by the product of an odd number of creation operators (e.g., what is known as a "fermionic" string).

Now we proceed to analyze the BCH expansion associated with Eq. (11):

$$\exp(-\hat{C})\hat{H}\exp(\hat{C}) = \hat{H} + \sum_k \frac{1}{k!}[\hat{H},\hat{C}]_k \quad (12)$$

We might be tempted to define, in analogy with the excitation CC case,[5] up, $s_A^+$, down, $s_A^-$, particle, $m_A$, and excitation, $s_A$, ranks for a string $A$ of creation and annihilation operators, as:

$$s_A^+ = \frac{1}{2}(n_v^c + n_o^a) \quad (13)$$

$$s_A^- = \frac{1}{2}(n_o^c + n_v^a) \quad (14)$$

$$m_A = s_A^+ + s_A^- \quad (15)$$

$$s_A = s_A^+ - s_A^- \quad (16)$$

Here, $n_o^c$ and $n_v^c$ ($n_o^a$ and $n_v^a$) are the number of occupied and virtual creation (annihilation) operators in $A$, respectively.

However, since in the creation-based CC ansatz we do not use the occupied/virtual separation, it is more natural to work with the following magnitudes:

$$s_A^c = \frac{1}{2}n^c \quad (17)$$

$$s_A^a = \frac{1}{2}n^a \quad (18)$$

$$s_A' = s_A^c - s_A^a \quad (19)$$

$$m_A = s_A^c + s_A^a \quad (20)$$

Notice that we are just counting the number of creation, $n^c$, and annihilation, $n^a$, operators in $A$. Since for the $\hat{\chi}_\mu$ operators $n^a = 0$, we will have:

$$m_{\chi_\mu} = s'_{\chi_\mu} = n_\mu \tag{21}$$

where $n_\mu$ is the number of creation operators in $\hat{\chi}_\mu$.

In Eq. (12) we will encounter the following nested commutators:

$$\hat{\Omega} = \left[...\left[\left[\hat{H}, \hat{\chi}_{\mu_1}\right], \hat{\chi}_{\mu_2}\right]..., \hat{\chi}_{\mu_k}\right] \tag{22}$$

Since we have restricted ourselves to have, at most, one fermionic string in the set of $\hat{\chi}_\mu$ operators, this means that we will have:

$$m_\Omega = m_H + \sum_{i=1}^{k} n_i - k \tag{23}$$

$$s'_\Omega = s'_H + \sum_{i=1}^{k} n_i \tag{24}$$

Therefore:

$$2s_\Omega^a = 2s_H^a - k \tag{25}$$

$$n_H^a \geq k \tag{26}$$

Finally, since the Hamiltonian has, at most, terms with 2 annihilation operators, this means that Eq. (12) will include, at most, double-nested commutators. In other words:

$$\exp(-\hat{C})\hat{H}\exp(\hat{C}) = \hat{H} + \left[\hat{H}, \hat{C}\right] + \frac{1}{2}\left[\left[\hat{H}, \hat{C}\right], \hat{C}\right] \tag{27}$$

The present analysis shows that a set of $\hat{\chi}_\mu$ operators must contain, at most a fermionic string for the creation-based CC to recover the characteristic multiplicatively-separable and product-form ansatz of the traditional excitation-based CC. This, in turn, implies that the BCH expansion of the similarity transformed Hamiltonian of creation-based

CC will not contain terms with three or more nested commutators. This provides new insights into the wavefunction forms that contain exponentials of creation operators, showing how we can simplify the work with these terms.

ACKNOWLEDGMENTS

We acknowledge support from the National Science Foundation CAREER award CHE-243986. Discussions with Paul W. Ayers and Taewon David Kim are gratefully acknowledged.